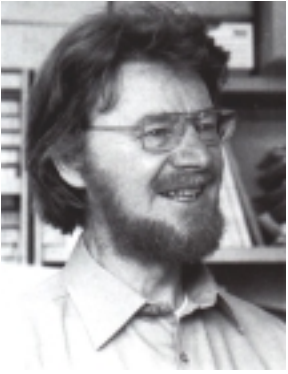

John Stewart Bell (1928-1990)

# Very promising hole in Bell's theorem

### R. Nakhmanson*

*Through the hole in Bell's theorem we can communicate with matter.*

"Imagination is more important than knowledge. Knowledge is limited. Imagination encircles the world."
 - Albert Einstein
   (1879-1955)

The J. S. Bell's theorem is based on the following assertion: if $P_a$ is a probability of result *a* measured on the particle **1** in the point **A** having a condition (e.g. angle of analyzer) $\alpha$, and $P_b$ is a probability of result *b* measured on the particle **2** in the distant point **B** having a condition $\beta$, then $\beta$ has no influence on the $P_a$, and vice versa. Here Bell and others saw the indispensable requirement of local realism and "separability". Mathematically it can be written as

$$P_{ab}(\lambda_{1i},\lambda_{2i},\alpha,\beta) = P_a(\lambda_{1i},\alpha) \times P_b(\lambda_{2i},\beta) \qquad \text{(Bell)}, \qquad (1)$$

where $P_{ab}$ is the probability of the joined result *ab*, and $\lambda_{1i}$ and $\lambda_{2i}$ are hidden parameters of particles **1** and **2** in an arbitrary local-realistic theory. Under the influence of Bell's theorem and the experiments following it and showing, that for entangled particles the condition (1) is no longer valid, some scientists reject locality. In this case an instantaneous action at a distance is possible, and one can write

$$P_{ab}(\lambda_{1i},\lambda_{2i},\alpha,\beta) = P_a(\lambda_{1i},\alpha,\beta) \times P_b(\lambda_{2i},\beta,\alpha) \qquad \text{(non-locality)}. \qquad (2)$$

In principle such a relation permits a description of any correlation between *a* and *b*, particularly correlation predicted by QM and observed in experiments. But in the frame of local realism the condition (1) is not indispensable. Instead, one can write

$$P_{ab}(\lambda_{1i},\lambda_{2i},\alpha,\beta) = P_a(\lambda_{1i},\alpha,\beta´) \times P_b(\lambda_{2i},\beta,\alpha´) \qquad \text{(forecast)}, \qquad (3)$$

where $\alpha´$ and $\beta´$ are the conditions of measurements in points **A** and **B**, respectively, as they can be forecast by particles at the moment of their parting. If the forecast is good enough, i.e., $\alpha´ \approx \alpha$ and $\beta´ \approx \beta$, then (3) practically coincides with (2) and has all its advantages plus locality.

Strictly speaking we suppose the particles possess some kind of consciousness. This idea has old tradition and seems very natural as compared with "many worlds" or "nonlocality". Such a "hidden parameter" missed by Bell can forecast future and provide EPR-correlation.

---

* Raoul Nakhmanson
  Frankfurt am Main, Germany
  E-mail: Nakhmanson@t-online.de

The idea leads to an alternative local-realistic interpretation of quantum mechanics (QM): In space there is a particle, its wave function is a product of its consciousness processing the known information about the world to optimize the particle's behavior. In other words, the wave function is not in the space but in the consciousness of the particle and is its strategy. This explains why it does not affect other particles being in the same 3d-space. If there are many particles, their distribution, e.g. interference fringes, looks like a product of real wave in real space.

If two or more particles have a common strategy they are "entangled" as long as they can forecast the future at the moment of their parting. The new-coming unlooked-for circumstances allow them gradually to cut off and forget the old partnership.

If the particle receives new information, it corrects the strategy, that is, the so-called "collapse" of wave function occurs, local and instant "FAPP". Wave function gives only a distribution of priorities. Taking this into account the particle makes its choice. The optimal tactics of proportional proving of all possibilities is randomization of this choice. Etc., including explanation of all QM-paradoxes.

Can this idea be proved? There are two possibilities:

   1. To destroy informational channels of particles. So, in 1982 Aspect et al. [1] in Paris switched analyzers very fast to cut off any subluminal informational contact between EPR-particles. They have confirmed the results of previous "static" experiments. Because of technical limitations they used periodical switching instead of a random one being desirable to prevent a forecast of the analyzer state. Sixteen years later Weihs et al. [2] in Innsbruck used the random generator - with the same result.

   These works, of course, were only first attempts in the complex, quaking, and provocative field where particles have much more experience as overweening "aggressors" impeding their life. For example, Weihs et al. wrote: "Selection of an analyzer direction has to be completely unpredictable, which necessitates a physical random number generator. A pseudo-random-number generator cannot be used, since its state at any time is predetermined" (p.5039). As a "physical random number generator" they used a "light-emitting diode illuminating a beamsplitter whose outputs are monitored by photomultipliers" (p.5041). But particles everyday contact with "physical" generators and nevertheless forecast future. If particles possess consciousness such "physical" generators are rather pseudo-random-number ones too. Perhaps a good "human" pseudo-random generator is for experiments preferable because it belongs to another civilization...

2. A more polite and interesting possibility is informational contact with particles, an attempt to speak with them. Fig. 1 introduces into the field. Fig. 1(*a*) shows a "black box" which is tested by the linearly polarized light beam. Inside of the box the beam meets a thick transparent glass plate fixed at the Brewster angle so that all photons pass the box. The glass plate manifest itself physically only by space shift $\Delta z$ and time delay $\Delta t$ of output photons. Further there is a movable mirror ("traffic divarication") which is controlled by the experimenter to turn or not to turn the beam. Such a control is a brutal one like a traffic barrier closing one of two branches of the road.

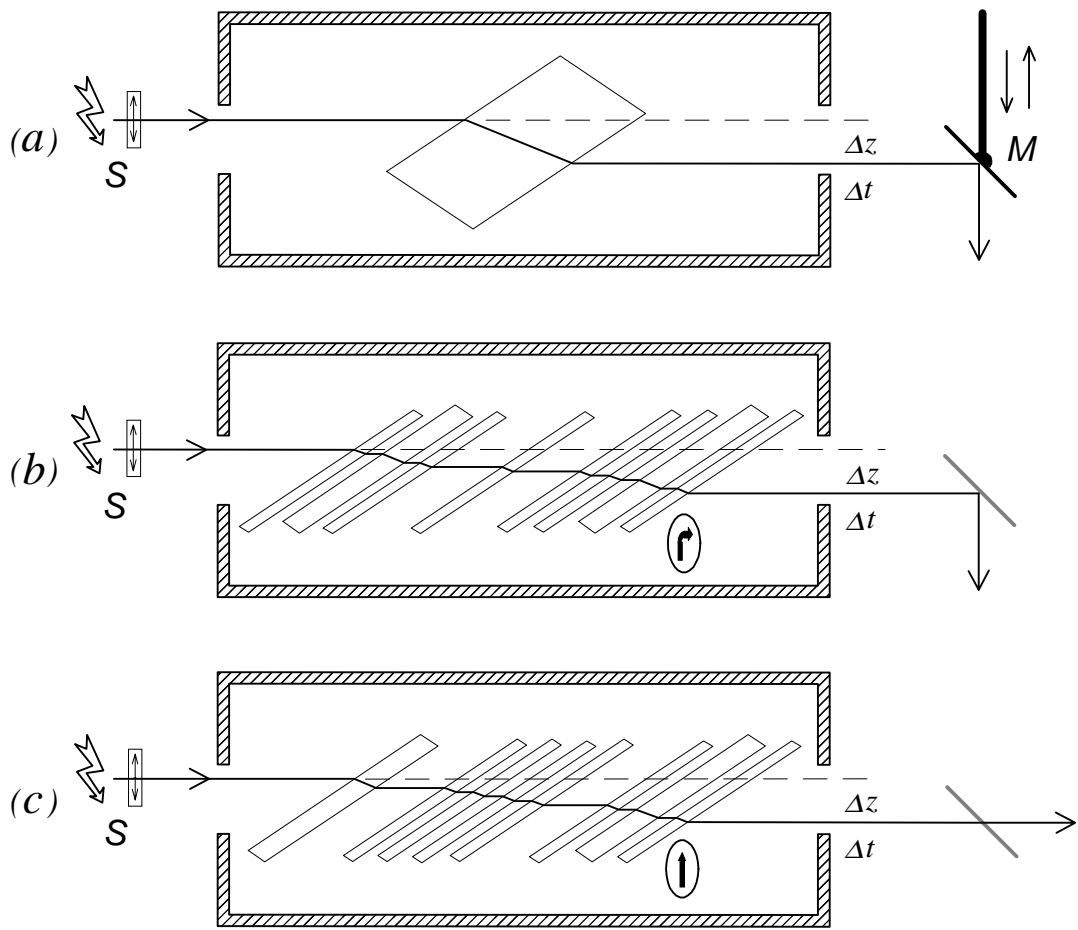

Fig. 1. (*a*) - "brutal" control, (*b*) and (*c*) - informational control .

In Figs. 1(*b*) and 1(*c*) the mirror is semi-transparent and immovable, and the thick glass plate is divided into eight thin plates, two of them being thicker than the remaining six. As before, the content of the black box manifests itself physically only by the same space shift $\Delta z$ and time delay $\Delta t$. But if the photons are intelligent and know English and Morse code, they can read the messages, namely

●—●  ●  ●●—●  = REF (reflect) in Fig. 1(*b*),

—  ●●●●  ●—●  = THR (through) in Fig. 1(*c*),

and follow the instructions. Such a control is an informational one like traffic signs on the road.

It is important to emphasize that the idea of informational experiments with particles, as it seems, has never been publicly discussed, and all experiments made with particles up to now cannot be considered as informational ones even in retrospect, that is, the revision of their results would not enable us to make any conclusion relating to this idea.

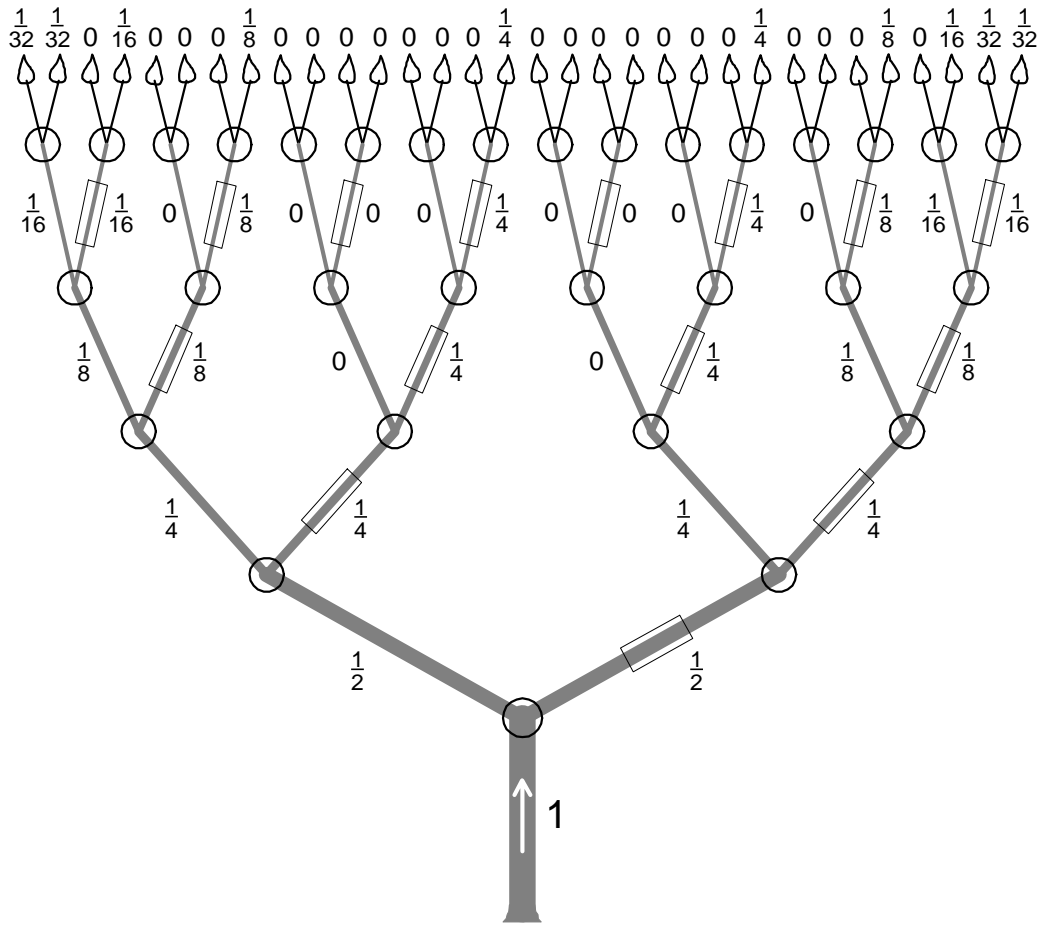

Fig. 2. Binary-tree experiment. Circles stand for beamsplitters, rectangles denote informational cells. Ciphers show the probability of detecting the particle in the case of the most rapid formation of a rigid, conservative conditioned reflex.

Fig. 2 shows the scheme of "binary-tree" experiment which does not require the particle's knowledge of English, Morse code, etc. The initial beam of microobjects (particles, atoms) enters into a system of beamsplitters (shown by circles). They can be semi-transparent mirrors for photons, crystals for electrons or neutrons, Stern-Gerlach apparatuses for atoms, etc. Fig. 2 shows only five floors of beamsplitters, but there can be as many as experimentally feasible. According to present-day theoretical ideas and practical experience, each of the output beams has the same intensity, namely, 1/32 of intensity of the initial beam (real beamsplitters may have, of course, some absorption, but here it is not a matter of principle).

Into each of the right branches of the binary tree (e.g. corresponding to reflection for photons) is introduced an "informational cell" (shown by rectangles), which is a device leaving unchanged the intensity of the beam passing it, but offering some information to particles. For polarized photons such a cell may be again a set of transparent plates fixed at the Brewster angle, and the information can be coded, say, by differences in the materials of plates, their thickness, and distances between them (as in Figs. 1(*b*), 1(*c*)) . The information in each subsequent floor is a sequel of that in the preceding floor. As a whole it can be music and some course of studies of language for following communication like ones being developed for the project "Search for Extra-Terrestrial Intelligence (SETI)".

The commonly accepted point of view is that the introduction of information cells will not change the uniform probability distribution of particles in the output beams. But if particles have consciousness, and are able to notice the information offered to them, they may become interested in it. After a number of floors, the particles should notice that the information is offered only in the right branches, and should prefer to choose their passing through the following beamsplitters. In other words, particles could develop a "conditioned reflex", of essentially the same kind as in behavior experiments on living beings.

Such an inquisitiveness of particles should lead to a change of their distribution in the output branches. For example, if the conditioned reflex appears immediately after comparison of right and left branches and particles have no interest to the left ones anymore, the distribution of probability to find the particle in different branches of the binary tree is like the one shown in Fig. 2 by ciphers.

Deviation from the uniform distribution of particles in the output beams will mean that the particles at least recognize the information offered and have an interest in it. Such an interest is thought to be an inherent attribute of each consciousness. This, however, still does not mean that the particles understand this information: people of modern times were interested in ancient hieroglyphic symbols long before they learned how to interpret them. To establish a deciphering stage, one can, starting from some floor of a binary tree, introduce some specific "requests" into the information cells. For example, one can "ask" particles to choose a left channel after the next beamsplitter rather than a right one. Because between the output branches of the binary tree and the trajectories of the particles there is a one-to-one interrelation, the honoring of such requests can easily be detected by an experimenter. However, the possibilities of an experiment typified in Fig. 2 are not exhausted by this. Purposefully choosing direction at each subsequent beamsplitter, the particle, in its turn, can send information to the experimenter using "right" and "left" as a binary code. For example, extreme left and extreme right trajectories in Fig. 2 present $00000 = 0$ and $11111 = 31$, respectively.

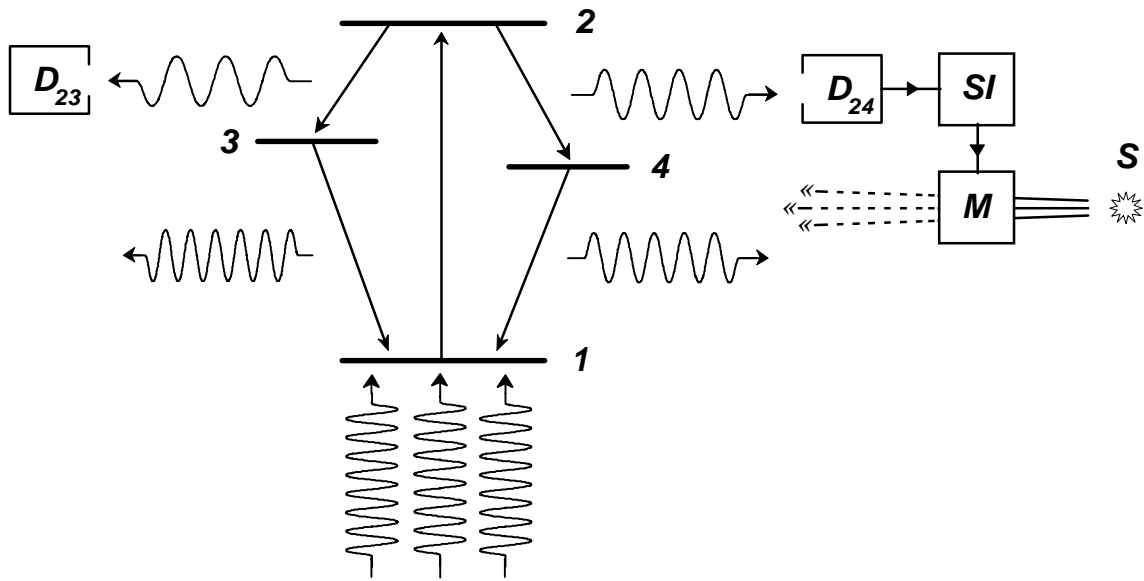

Fig. 3  Informational experiment with a single atom. 1, 2, 3, and 4 are the energy levels; $D_{23}$ and $D_{24}$ are detectors, $S$ is the source of light, $M$ is the modulator, $SI$ is the source of information .

The experiments illustrated in Fig. 1 and Fig. 2 can be called "coordinate-impulse" ones to distinguish them from the "energy-time" experiment whose scheme is shown in Fig. 3 . Here a four-level quantum system, e.g., an atom, with one low (1), one high (2), and two intermediate (3,4) energy levels is pumped by intensive radiation inducing the 1→2 transition, so that the atom does not stay in the state 1 but immediately is translated into the state 2. From it, the atom makes a transition spontaneously to the state 3 or 4, and later makes a transition to the state 1 completing the cycle. The radiation corresponding to some of the transitions 2→3, 2→4, 3→1, and 4→1 are detected (in Fig. 3 two detectors are shown). Besides, there is an informational action on the atom, e.g., by modulation of light coming from the source $S$ . The modulator $M$ is controlled by the source of information $SI$, which, in turn, is connected with one or more detectors to close the feedback loop.

The feedback works in such a way as to stimulate a channel and rate of transitions, in the case of Fig. 3, the 2→4→1 transitions. The source $SI$ sends a message, e.g., one line of a page or a measure of a music, only if it receives a signal from detector $D_{24}$. Each next message continues the previous one, i.e., is the next line or the next measure.

If the atom has a consciousness and is interested in the information being proposed, it develops a conditioned reflex and will prefer the 2→4 transition to the 2→3 one. Besides, the rates of both 2→4 and 4→1 transitions must increase. All this can be registered by the experimenter. To be sure that the effect is connected with information, one can make a control experiment to cut off the feedback or/and to use some "trivial" information, etc.

Like with the scheme of Fig. 2, in the last case one may hope to observe not only an interest of a quantum object to receive a new information, but deciphering it also, as well as the sending of messages from the object to the experimenter being coded in states of the atom and time intervals between the states.

In 1876 A. G. Bell invented telephone for communication between people. Eighty five years later J. S. Bell left a hole in his theorem to give us possibility to communicate with matter.

**References**

[1] A. Aspect, J. Dalibard, and G. Roger, Phys. Rev. Lett. **49**, 1804 (1982).

[2] G. Weihs, T. Jennewein, Ch. Simon, H. Weinfurter, and A. Zeilinger, Phys. Rev. Lett. **81**, 5039 (1998).

See also: http://xxx.lanl.gov/pdf/physics/0004047 and
http://xxx.lanl.gov/pdf/physics/0005042 .

*Note:* This poster was submitted for the conference in commemoration of John S. Bell "Quantum [Un]speakables", Vienna, Austria, 10-14 Nov. 2000